# Direct visualization of percolating metal-insulator transition in V$_2$O$_3$ using scanning microwave impedance microscopy

Weiyan Lin[1], Huanyu Zhang[2], Yoav Kalcheim[3,4], Xinchen Zhou[2,5], Fubao Yang[2,5], Yang Shi[1], Yang Feng[2,6], Yihua Wang[2,7], Jiping Huang[2,5], Ivan K. Schuller[3,4], Xiaodong Zhou[1,9,10†], and Jian Shen[1,2,7,8,9,10†]

[1]State Key Laboratory of Surface Physics and Institute for Nanoelectronic Devices and Quantum Computing, Fudan University, Shanghai, China.
[2]Department of Physics, Fudan University, Shanghai, China.
[3]Department of Physics, University of California San Diego, La Jolla, California 92093, USA
[4]Center for Advanced Nanoscience, University of California San Diego, La Jolla, California 92093, USA
[5]Key Laboratory of Micro and Nano Photonic Structures (MOE), Fudan University, Shanghai, China
[6]Beijing Academy of Quantum Information Science, Beijing, China
[7]Shanghai Research Center for Quantum Sciences, Shanghai, China.
[8]Collaborative Innovation Center of Advanced Microstructures, Nanjing, China
[9]Shanghai Qi Zhi Institute, Shanghai, China
[10]Zhangjiang Fudan International Innovation Center, Fudan University, Shanghai, China

†Emails: zhouxd@fudan.edu.cn, shenj5494@fudan.edu.cn

Abstract

Using the extensively studied V$_2$O$_3$ as a prototype system, we investigate the role of percolation in metal-insulator transition (MIT). We apply scanning microwave impedance microscopy to directly determine the metallic phase fraction $p$ and relate it to the macroscopic conductance $G$, which shows a sudden jump when $p$ reaches the percolation threshold. Interestingly, the conductance $G$ exhibits a hysteretic behavior against $p$, suggesting two different percolating processes upon cooling and warming. Based on our image analysis and model simulation, we ascribe such hysteretic behavior to different domain nucleation and growth processes between cooling and warming, which is likely caused by the decoupled structural and electronic transitions in $V_2O_3$ during MIT. Our work provides a microscopic view of how the interplay of structural and electronic degrees of freedom affects MIT in strongly correlated systems.

## I. INTRODUCTION

Strongly correlated systems are known to often exhibit a strong spatial inhomogeneity and showcase different electronic phases on length scales from micro- to nanometers[1]. Their macroscopic physical properties and related phase transitions depend sensitively on actual spatial distribution of the electronic phase domains. In particular, metal-insulator transitions (MIT), while being ultimately related to how electrons self-organize themselves in real space[2-4], is often described by a percolation theory given that both metallic and insulating phases coexist during the transition. As a simple model, percolation theory offers a unified perspective to understand the critical behavior of MIT in different systems by performing a scaling analysis even though the underlying mechanism of the transition varies[5-7]. Under percolation theory, the resistivity behavior during MIT can be understood without knowing its microscopic mechanism.

Although conceptually appealing, the percolation theory of MIT has rarely received experimental scrutiny at a quantitative level albeit with some general arguments in previous works. What is essential for such an analysis is to directly relate a metallic phase fraction $p$ to the global conductance $G$ of the system. While such $G(p)$ can be readily calculated in percolation models, its experimental determination is not straightforward because the metallic phase fraction $p$ is often guised in MIT experiments by an actual control parameter, such as the temperature $T$, the applied magnetic field $B$ or carrier density $n$. One has to independently extract $p$ and $G$ from experiments to obtain such $G(p)$ and compare with models. That requires the involvement of both global transport and microscopic characterizations.

In this work, we use $V_2O_3$ as a model system to investigate its thermally driven MIT from a perspective of percolation. We directly visualize MIT in $V_2O_3$ using scanning microwave impedance microscopy (sMIM), which allows us to determine the metallic phase fraction $p$ during the transition and relate it to the macroscopic conductance $G$ of the system. The overall shape of $G(p)$ is typical of a percolating process especially for the regime near the percolation threshold. However, a sizable thermal hysteresis of $G(p)$ develops in the regime away from the percolation threshold. The contrasting behavior between cooling and warming indicates there exist two different percolating processes dictated by the domain wall energy in the formation of new domains. Such domain wall energy difference is closely tied to the interplay of structural and electronic transitions in $V_2O_3$.

## II. RESULTS AND DISCUSSION

$V_2O_3$ is a strongly correlated system with pronounced MIT[8,9]. It changes from a high-temperature paramagnetic metal to a low-temperature antiferromagnetic insulator ($T_c \sim 160$ K) with the resistivity jump reaching a factor of $10^6$. This transition is accompanied by a structural change from the high-temperature corundum structure to the low-temperature monoclinic structure. Further study shows that MIT in $V_2O_3$ can be classified as a Mott transition occurring together with structural and magnetic transitions[10,11]. $V_2O_3$ also displays a current-/voltage-induced resistive-switching and thus finds its application in the emerging field of neuromorphic computing[13]. Such resistive-switching is intimately related to the thermally driven MIT of $V_2O_3$[14].

The MIT of $V_2O_3$ is also considered as a percolative type. The coexistence of metallic and insulating phases during the transition has been revealed by various experimental probes[14-18]. Although they generally claim the percolation at play in

MIT, the quantitative analysis is quite limited. Here we combine both transport and local imaging technique, together with model simulations, to conduct a quantitative analysis of percolation. Our $V_2O_3$ sample is a 300 nm thick film grown on a r-cut sapphire substrate by radiofrequency magnetron sputtering from a $V_2O_3$ target[19]. Transport measurement in Fig. 1(a) shows that the resistance of the film changes by 5 orders of magnitude during the temperature driven MIT with a strong thermal hysteresis. Such hysteresis is due to the first-order phase transition and indicates phase coexistence during the transition from one end-phase to the other. We employ sMIM to directly visualize such phase coexistence in real space. Figure 1(b) shows the schematic of sMIM experiment. A 3 GHz microwave is delivered to an atomic force microscope tip apex and the reflected microwave signal is collected and demodulated into two output signals sMIM-Im and sMIM-Re, which are proportional to the imaginary and real components of the tip-sample admittance, respectively. This tip-sample admittance is solely determined by the spatial profile of the complex dielectric constant $\hat{\varepsilon}(r) = \varepsilon'(r) + i[\varepsilon''(r) + \sigma(r)/\omega]$ for the entire region around the tip-sample interface. For systems with MIT, it is the sample's local variation of conductivity $\sigma(r)$ that affects the tip-sample admittance and gives rise to the sMIM imaging contrast[20]. Therefore, we take advantage of sMIM's high spatial resolution (~ 100 nm) and electrical sensitivity to the local conductivity to identify insulating and metallic phases in $V_2O_3$. To be specific, the sMIM-Im signal is used throughout the work as it increases monotonically with the local conductivity.

Figure 1(c) shows a series of selected sMIM-Im images taken at the same area during the MIT under the cooling and warming process. These images are normalized to a gold electrode to compensate the temperature induced sMIM signal background shift (see supplementary materials S1). Therefore, the maximum sMIM signal value in the normalized image of Fig. 1(c) is zero as denoted in the color scale. We apply a color scheme in which yellow (blue) represents high (low) sMIM signal corresponding to metallic (insulating) phase. As shown in Fig. 1(c), the system starts from a uniform metallic state at high temperature (185 K). At lower temperatures, isolated insulating phases emerge (173 K), grow (167 K-164 K) and overtaken metallic phases (161 K). The system finally reaches a uniform insulating state at 155 K. The whole process reverses on heating at this point. Metallic phases appear (164 K), expand (167-170 K) and fragment insulating phases (176 K). The system returns back to a uniform metallic state at 185 K. Consistent with transport, sMIM images show a thermal hysteresis as well, i.e., the metallic phase reaches the same fraction during warming at temperature 3 K higher than that during cooling. The domain geometry of phase coexistence develops a bi-directional striped pattern similar to previous reports[14,17]. These types of patterns are attributed to elastic energy minimization as strain changes due to the structural transition, which was found to be robustly coupled to the electronic transition in $V_2O_3$[18].

The coupling of electronic orders to external factors in our sample can be inferred from another experiment. As shown in Fig. 2(a-b), we compare sMIM images taken at the same temperatures ($T =$167 K, 164 K, 161 K) in the cooling processes for two consecutive thermal cycles of MIT. The domain spatial distribution is highly reproducible, i.e., almost all the metallic phases reappear at the same spatial locations after a thermal cycle. This is confirmed by a two-dimensional (2D) cross-correlation analysis for each temperature. As shown in Fig. 2(c), the 2D cross-correlation maps all exhibit peaks in the center, indicating that the domain patterns were well aligned to the same locations. We extract the maximum cross-correlation coefficients from these images to have $r_{xy} =$0.45, 0.49 and 0.28 for 167 K, 164 K and 161 K, respectively. These values suggest a relatively strong correlation between the sMIM images measured at different thermal cycles, especially when the system approaches to the transition point. This observation indicates a strong domain pinning effect during nucleation and growth. Similar pinning effect of domain configuration has been reported in manganites[21].

Our spatial imaging of phase coexistence in $V_2O_3$ allows us to conduct a more quantitative analysis of the MIT with a view of percolation. We directly determine the metallic phase fraction $p$ from sMIM images and show it as a function of temperature $p(T)$ in Fig. 3(a). Several features appear in the plot. $p(T)$ shows a rapid increase at lower temperatures, and slows down when $p \sim$ 50%-60% before it takes on another rise-up until 100%. A thermal hysteresis is identified between cooling and warming curves. However, such hysteresis is larger at lower temperatures than that at higher temperatures, i.e., two $p(T)$ curves are not parallel with each other in Fig. 3(a). This gives us a first hint that cooling and warming processes are different albeit with a thermal lag. To better appreciate the consequence of such $p(T)$ behavior, we plot $G(p)$ in Fig. 3(b) in which both $G$ and $p$ are independently determined at the same temperature. Note that we put $p$ in the vertical axis so it can be directly compared to Fig. 3(a). As discussed in the introduction, $G(p)$ encodes all essential charterers of MIT that one wants to closely compare with the percolation model. Some key features are revealed here. First, for both cooling and warming process, $G(p)$ undergoes a sudden increase when $p$ is above 50% as reveled by the change of the slope. This indicates a percolating process and 50% can be defined as a percolation threshold above which the metallic phase interconnects the whole system resulting in a jump of the conductance. In standard percolation theory, 50% percolation threshold corresponds to bond percolation model in 2D square lattice or site percolation model in 2D triangular lattice[5]. Second, $G(p)$ of cooling and warming collapses into a single curve near the percolation threshold when $p \sim$ 50%-60% , but displays a visible thermal hysteresis away from it, especially for the regime above the threshold. This uncovers a fact that cooling and warming processes are two different percolating processes with varied $G(p)$ behavior as elaborated below.

A major assumption in percolation model is that the metallic and insulating phases can be treated on an equal footing. That is, adding metallic regions to an insulating host is equivalent to adding insulating regions to a metallic host as long as they reach the same metallic fraction. This is particularly true when analyzing the critical phenomena of MIT close to the threshold $G(p) \propto |p - p_c|^t$, where what matter is the distance to $p_c$ not the approaching direction. For instance, in a cooling process what really happens in the system is that an insulating phase nucleates and grows from a metallic host. Under percolation model, it can be treated as an equivalent warming process in which a metallic phase emerges from an insulating host. As shown in Fig. 3(b), the system indeed behaves like that, so the $G(p)$ of cooling looks very similar to that of warming especially in the regime near percolation threshold. This observation is another evidence that MIT of $V_2O_3$ displays a strong percolating character. However, the thermal hysteresis of $G(p)$ away from the percolation threshold indicates that the MIT of cooling and warming represents two different percolation processes, rather than a single process with different approaching directions. The varied line-shape of $G(p)$ between cooling and warming suggests that the metallic phase experiences different domain growth procedures.

To uncover the factors causing such a difference, we analyze the sMIM images with the view that the MIT can be considered as a process in which a minority phase nucleates and grows from a majority phase, and the sMIM imaging snapshots the domain configuration at each step of the domain growth. Here the minority phase corresponds to the insulating (metallic) phase in the cooling (warming) process. As shown in Fig. 4(a), the first information we obtained from such image analysis is the number of domains and/or nucleation sites of the minority phase versus its area fraction for cooling and warming processes. The error bar reflects the uncertainty of such domains and/or nucleation sites determination from sMIM images (see supplementary material S2). Figure 4(a) shows that this number of both processes quickly rises up and peaks at $p \sim$ 20% before gradually going down. This is understandable because domain nucleation and proliferation certainly dominates at the beginning. As the domains grow and coalesce to form bigger ones, the number of domains is reduced. Figure 4(a) also shows that for a given area fraction, one has more domains and/or nucleation sites in the warming than the cooling, especially at the beginning of nucleation. The second information obtained from our analysis is that the domain growth of minority phase is a net balance of new areas gained and old areas lost (see supplementary materials S3). Phase competition happens everywhere inside the sample determining its metallicity at a given temperature.

Based on our image analysis, we now run a percolation model simulation to see if we can reproduce the main experimental results. We choose a random-resistor network model, i.e., a lattice model of percolation theory for analyzing transport properties[22], which has been previously applied to study MIT in both $V_2O_3$ and other strongly correlated systems[14,23]. In particular, we use a site percolation model in 2D square lattice to simulate the domain nucleation and growth (see supplementary materials S3) for both cooling and warming processes, where the minority phase corresponds to insulating and metallic phase, respectively. Two important factors are considered in the simulation based on our image analysis above. First, the number of nucleation sites is more in warming than in cooling, with the number ratio between warming and cooling being an input parameter in the simulation. Second, we simulate the minority phase domain growth by considering both the area expansion and retraction. At each step of the simulation, there is a certain probability for an area to become metallic or insulating, which reflects the underlying domain growth mechanism. Such probability is another input parameter. Figure 4(b) shows the simulated metallic area fraction versus simulation steps which can be directly compared to $p(T)$ in Fig. 3(a). Note that the area fraction of the metallic phase in the cooling process is obtained by simulating the insulating phase growth at each step which is then subtracted from 1. We can also calculate the conductance of such random-resistor network for a given domain configuration (see supplementary materials S4). For such calculation, we need to assign a resistance ratio between the insulating and metallic phase. Figure 4(c) shows the calculated normalized conductance as a function of metallic phase fraction which can be compared to $G(p)$ in Fig. 3(b).

Our model simulation generally reproduces the main features of our experimental results, i.e., the different line-shapes of $p(T)$ and $G(p)$ between cooling and warming. For domain growth, we see a large thermal hysteresis at the beginning in Fig. 4(b) which shrinks and disappears later, closely resembling the $p(T)$ behavior in Fig. 3(a). The calculated $G(p)$ in Fig. 4(c) also qualitatively resembles Fig. 3(b) in the sense that conductance difference appears when $p$ passes a percolation threshold. $G(p)$ in Fig. 4(c) can be understood in the following sense: cooling process has an effective lower percolation threshold, so its conductance rises up earlier than warming leading to the conductance gap above the percolation threshold. What is missing in such simulation is the "slow down" of domain growth near the percolation threshold seen in $p(T)$ of Fig. 3(a). This may be caused by the crossover from percolation to cluster regime, which has been proposed in polymer mixtures with phase separation[24].

The physical origin of the difference between cooling and warming in the percolating process may be caused by the decoupling of the structural and electronic transition in MIT. A recent near-field infrared imaging, combined with x-ray diffraction, reveals a 6 K thermal offset between electronic and structural transitions by identifying an intermediate monoclinic metallic phase during the transition[17]. More evidence for a structural transition preceding the MIT came from the surface acoustic wave measurements[25] and Raman spectroscopy[26]. This decoupling of structural and electronic transition offers a possible explanation for our experiments. In conventional domain theory, new domain emerges to save the system's total free energy at the cost of domain wall energy. We ascribe the domain wall energy in $V_2O_3$ to an elastic energy at the boundary of two structures. Therefore, the fact that we see more domains in the warming process indicates a negligible domain wall energy due to the transition from a monoclinic insulating phase to a monoclinic metallic phase at lower temperature without structural change. On the other hand, the system costs more domain wall energy in the cooling when a monoclinic insulating phase grows out of a corundum metallic host leading to fewer domains.

### III. CONCLUSION

In summary, we apply sMIM to directly visualize the percolating process of MIT in $V_2O_3$. We identify different percolating processes for MIT in the warming and cooling as evidenced by their different domain evolution. Such difference is dictated by the domain wall energy which is closely related to the interplay of structural and electronic transitions in $V_2O_3$. Our work shows that it is critical to correlate macroscopic transport properties to microscopic processes to understand the percolative MIT in strongly correlated materials.

The work at Fudan University is supported by National Natural Science Foundation of China (Grant Nos. 12074080, 11804052, 11827805, 11725521 and 12035004), National Postdoctoral Program for Innovative Talents (Grant No. BX20180079), Shanghai Science and Technology Committee Rising-Star Program (19QA1401000), The Science and Technology Commission of Shanghai Municipality (Grant No. 20JC1414700), Major Project (Grant No. 2019SHZDZX01) and Ministry of Science and Technology of China (Grant Nos. 2017YFA0303000 and 2021YFA1400100). The work (synthesis, structural characterization and global transport of $V_2O_3$) at UCSD is funded by the US Air Force Office of Scientific Research (award No. FA9550-20-1-0242).

**References**

[1] K. A. Moler, Nat. Mater. **16**, 1049 (2017).
[2] M. Imada, A. Fujimori, and Y. Tokura, Rev. Mod. Phys. **70**, 1039 (1998).
[3] N. F. Mott, *Metal-insulator transition* (Taylor & Francis, London, 1990).
[4] E. Abrahams, P. W. Anderson, D. C. Licciardello, and T. V. Ramakrishnan, Phys. Rev. Lett. **42**, 673 (1979).
[5] D. Stauffer and A. Aharony, *Introduction To Percolation Theory* (Taylor & Francis, London, 1992).
[6] P. Limelette, A. Georges, D. Jerome, P. Wzietek, P. Metcalf, and J. M. Honig, Science **302**, 89 (2003).
[7] V. Dobrosavljevic, N. Trivedi, and J. M. V. Jr., *Conductor Insulator Quantum Phase Transitions* (Oxford University Press, 2012).
[8] D. B. McWhan, A. Menth, J. P. Remeika, W. F. Brinkman, and T. M. Rice, Phys. Rev. B **7**, 1920 (1973).
[9] D. B. McWhan, A. Jayaraman, J. P. Remeika, and T. M. Rice, Physical Review Letters **34**, 547 (1975).
[10] B. A. Frandsen *et al.*, Phys. Rev. B **100**, 235136 (2019).

[11] J. Trastoy *et al.*, Phys. Rev. B **101**, 245109 (2020).
[12] S. Biermann, A. Poteryaev, A. I. Lichtenstein, and A. Georges, Phys. Rev. Lett. **94**, 026404 (2005).
[13] J. del Valle *et al.*, Nature **569**, 388 (2019).
[14] Matthias Lange, Stefan Guénon, Yoav Kalcheim, Theodor Luibrand, Nicolas Manuel Vargas, Dennis Schwebius, Reinhold Kleiner, Ivan K. Schuller, and D. Koelle, arXiv:2009.12536 (2020).
[15] S. Guénon, S. Scharinger, S. Wang, J. G. Ramírez, D. Koelle, R. Kleiner, and I. K. Schuller, EPL **101**, 57003 (2013).
[16] M. M. Qazilbash *et al.*, Phys. Rev. B **83**, 165108 (2011).
[17] A. S. McLeod *et al.*, Nat. Phys. **13**, 80 (2017).
[18] Y. Kalcheim, N. Butakov, N. M. Vargas, M.-H. Lee, J. del Valle, J. Trastoy, P. Salev, J. Schuller, and I. K. Schuller, Phys. Rev. Lett. **122**, 057601 (2019).
[19] I. Valmianski, J. G. Ramirez, C. Urban, X. Batlle, and I. K. Schuller, Phys. Rev. B **95**, 155132 (2017).
[20] Z. D. Chu, L. Zheng, and K. J. Lai, Annu. Rev. Mater. Res. **50**, 105 (2020).
[21] K. J. Lai, M. Nakamura, W. Kundhikanjana, M. Kawasaki, Y. Tokura, M. A. Kelly, and Z. X. Shen, Science **329**, 190 (2010).
[22] S. Kirkpatrick, Rev. Mod. Phys. **45**, 574 (1973).
[23] M. Mayr, A. Moreo, J. A. Verges, J. Arispe, A. Feiguin, and E. Dagotto, Phys. Rev. Lett. **86**, 135 (2001).
[24] H. Takeno and T. Hashimoto, J. Chem. Phys. **107**, 1634 (1997).
[25] J. Kundel *et al.*, Appl. Phys. Lett. **102** (2013).
[26] S. S. Majid, D. K. Shukla, F. Rahman, K. Gautam, R. J. Choudhary, V. G. Sathe, and D. M. Phase, Appl. Phys. Lett. **110** (2017).

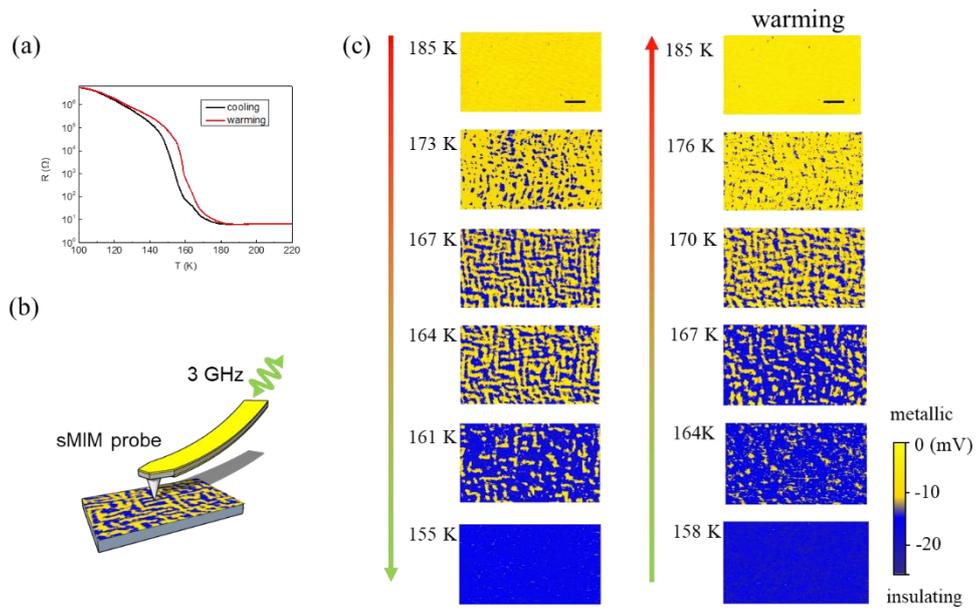

FIG. 1. (a) Temperature dependence of the film resistance for a cooling and warming cycle. (b) The schematic of sMIM experiment. (c) Temperature dependent sMIM images for a cooling and warming cycle of MIT. Scale bar is 5μm.

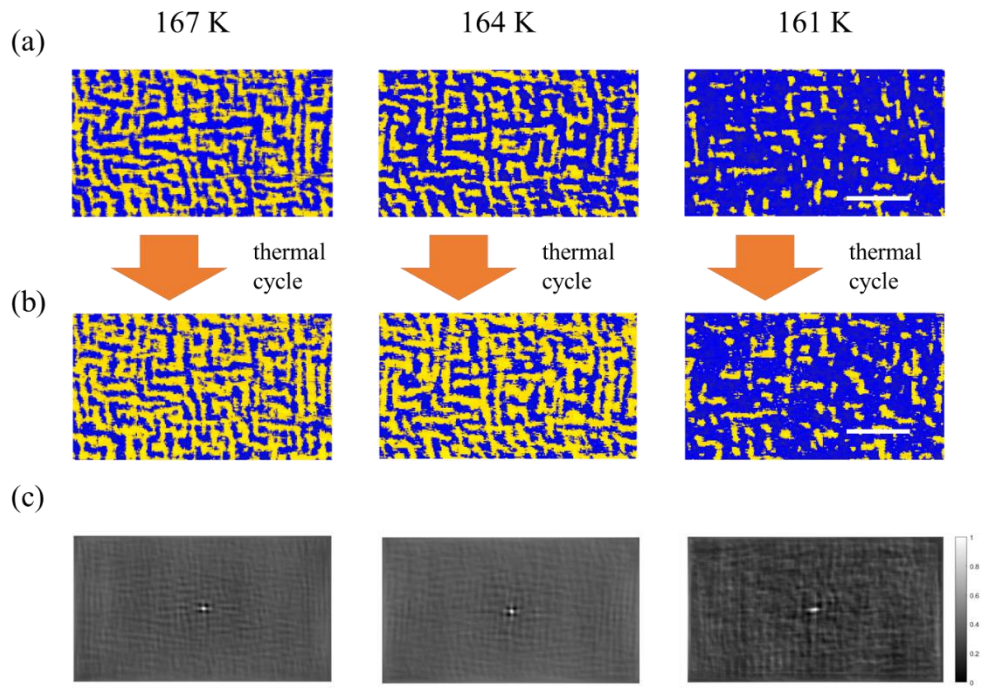

FIG. 2. (a),(b) sMIM images during the cooling process of the (a) first and (b) second thermal cycle at 167 K, 164 K and 161 K. (c) The 2D cross-correlation map between the sMIM images in (a) and (b). Scale bar is 5μm.

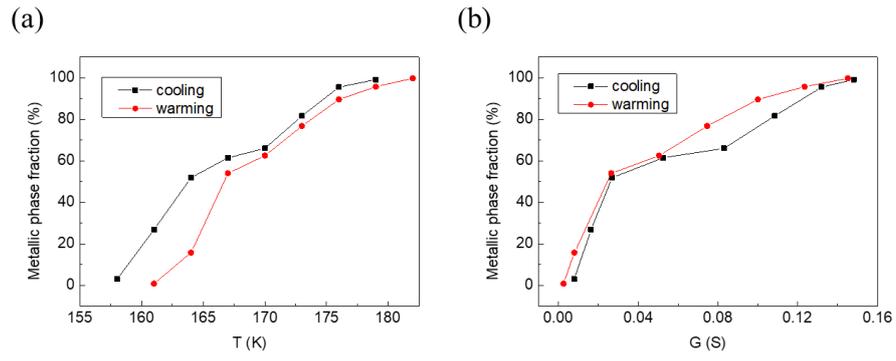

FIG. 3. (a) Temperature dependent metallic phase fraction. (b) The global conductance as a function of metallic phase fraction.

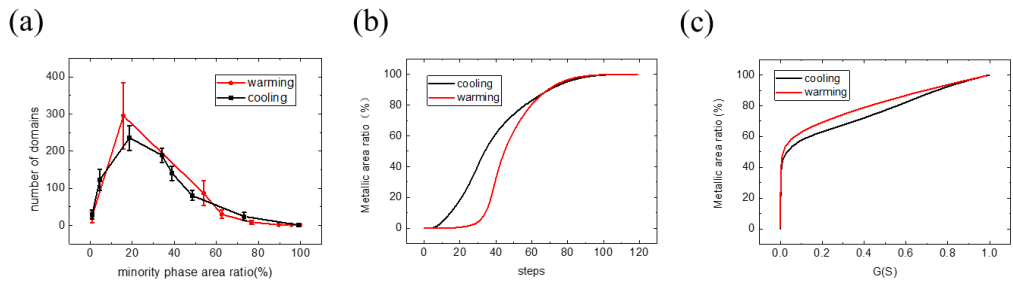

FIG. 4. (a) The number of the minority phase domains and/or nucleation sites versus its area fraction for cooling and warming. (b) Simulated domain growth for cooling and warming. (c) Calculated global conductance as a function of metallic phase fraction for cooling and warming.